\documentclass{article}

\usepackage{arxiv}

\usepackage[utf8]{inputenc} 
\usepackage[T1]{fontenc}    
\usepackage{hyperref}       
\usepackage{url}            
\usepackage{booktabs}       
\usepackage{amsfonts}       
\usepackage{nicefrac}       
\usepackage{microtype}      
\usepackage{lipsum}		    
\usepackage{graphicx}
\usepackage{subcaption}
\usepackage{tabularx}
\usepackage[numbers]{natbib}
\usepackage{doi}
\usepackage{caption}
\usepackage{threeparttable}
\usepackage{makecell} 
\usepackage{multirow}
\usepackage{placeins}
\usepackage{float} 
\usepackage[affil-it]{authblk} 
\usepackage{hyperref} 

\usepackage[export]{adjustbox}

\title{MRSegmentator: Multi-Modality Segmentation of 40 Classes in MRI and CT}

\author[1,2]{\textbf{Hartmut Häntze} \textsuperscript{*}}
\author[1]{\textbf{Lina Xu} \textsuperscript{*}}
\author[3]{Christian J. Mertens}
\author[1,4]{Felix J. Dorfner}
\author[1]{Leonhard Donle}
\author[3]{Felix Busch}
\author[3]{Avan Kader}
\author[3]{Sebastian Ziegelmayer}
\author[5]{Nadine Bayerl}
\author[6]{Nassir Navab}
\author[7,8,9]{Daniel Rueckert}
\author[9]{Julia Schnabel}
\author[10,11,12]{Hugo JWL Aerts}
\author[13]{Daniel Truhn}
\author[14]{Fabian Bamberg}
\author[14]{Jakob Weiß}
\author[14]{Christopher L. Schlett}
\author[14]{Steffen Ringhof}
\author[15]{Thoralf Niendorf}
\author[15]{Tobias Pischon}
\author[16]{Hans-Ulrich Kauczor}
\author[16]{Tobias Nonnenmacher}
\author[17]{Thomas Kröncke}
\author[18]{Henry Völzke}
\author[19]{Jeanette Schulz-Menger}
\author[20]{Klaus Maier-Hein}
\author[2]{Mathias Prokop}
\author[2]{Bram van Ginneken}
\author[2]{Alessa Hering}
\author[3]{Markus R. Makowski}
\author[3]{\textbf{Lisa C. Adams} \textsuperscript{$\dagger$}}
\author[3,21]{\textbf{Keno K. Bressem} \textsuperscript{$\dagger$}\textsuperscript{$\ddagger$}}

\affil[1]{Department of Radiology, Charité - Universitätsmedizin Berlin corporate member of Freie Universität Berlin and Humboldt Universität zu Berlin, Germany}
\affil[2]{Diagnostic Image Analysis Group, Radboud University Medical Center, The Netherlands}
\affil[3]{Department of Diagnostic and Interventional Radiology, School of Medicine and Health, Klinikum rechts der Isar, TUM University Hospital, Germany}
\affil[4]{Athinoula A. Martinos Center for Biomedical Imaging, Massachusetts General Hospital and Harvard Medical School, USA}
\affil[5]{Institute of Radiology, University Hospital Erlangen, Friedrich-Alexander-Universität (FAU) Erlangen-Nürnberg, Germany}
\affil[6]{Laboratory for Computer Aided Medical Procedures, Technical University of Munich, Germany}
\affil[7]{Chair for AI in Medicine and Healthcare, Klinikum rechts der Isar, Technical University Munich, Germany}
\affil[8]{Department of Computing, Imperial College London, UK}
\affil[9]{Institute for Advanced Study, Technical University Munich, Germany}
\affil[10]{Artificial Intelligence in Medicine (AIM) Program, Mass General Brigham, Harvard Medical School, USA}
\affil[11]{Departments of Radiation Oncology and Radiology, Dana-Farber Cancer Institute and Brigham and Women’s Hospital, USA}
\affil[12]{Radiology and Nuclear Medicine, CARIM \& GROW, Maastricht University, Netherlands}
\affil[13]{Department of Diagnostic and Interventional Radiology, University Hospital Aachen, Germany}
\affil[14]{Department of Diagnostic and Interventional Radiology, Medical Center - University of Freiburg, Faculty of Medicine, University of Freiburg, Germany}
\affil[15]{Berlin Ultrahigh Field Facility (B.U.F.F.), Max Delbrück Center for Molecular Medicine in the Helmholtz Association, Germany}
\affil[16]{Clinic for Diagnostic and Interventional Radiology, Heidelberg University Hospital, Germany}
\affil[17]{Department of Diagnostic and Interventional Radiology and Neuroradiology, Universitätsklinikum Augsburg, Germany}
\affil[18]{Institute for Community Medicine, University Medicine Greifswald, Germany}
\affil[19]{Experimental Clinical Research Center, Charité - Universitätsmedizin Berlin corporate member of Freie Universität Berlin and Humboldt Universität zu Berlin, Germany}
\affil[20]{Division of Medical Image Computing, Deutsches Krebsforschungszentrum Heidelberg, Germany}
\affil[21]{Department of Cardiovascular Radiology and Nuclear Medicine, School of Medicine and Health, German Heart Center, TUM University Hospital, Technical University of Munich, Germany}

\renewcommand{\headeright}{}
\renewcommand{\shorttitle}{MRSegmentator}

\hypersetup{
pdftitle={MRSegmentator: Multi-Modality Segmentation of 40 Classes in MRI and CT},
pdfsubject={MRI, Segmentation, Cross-Modality, Artificial Intelligence, Radiology},
pdfauthor={Hartmut Häntze, Lina Xu, Christian J. Mertens, Felix J. Dorfner, Leonhard Donle, Felix Busch, Avan Kader, Sebastian Ziegelmayer, Nadine Bayerl, Nassir Navab, Daniel Rueckert, Julia Schnabel, Hugo JWL Aerts, Daniel Truhn, Fabian Bamberg, Jakob Weiß, Christopher L. Schlett, Steffen Ringhof, Thoralf Niendorf, Tobias Pischon, Hans-Ulrich Kauczor, Tobias Nonnenmacher, Thomas Kröncke, Henry Völzke, Jeanette Schulz-Menger, Klaus Maier-Hein, Marcus R. Makowski, Mathias Prokop, Bram van Ginneken, Alessa Hering, Lisa C. Adams, Keno K. Bressem},
pdfkeywords={MRI, Segmentation, Artificial Intelligence, Radiology},
}

\begin{document}
\maketitle
\textsuperscript{*}These authors contributed equally as first authors.\\
\textsuperscript{$\dagger$}These authors contributed equally as last authors.\\
\textsuperscript{$\ddagger$}Corresponding author: keno.bressem@tum.de \\

\newpage
\abstract{
    \textbf{Background:} Automated MRI analysis is limited by low resolution, lack of standardized signal intensity values, and variability in acquisition protocols.
    
    \textbf{Purpose:} To develop and evaluate a deep learning model for multi-organ segmentation of MRI scans.
    
    \textbf{Materials and Methods:} 
    The model was trained on 1,200 manually annotated 3D axial MRI scans from the UK Biobank, 221 in-house MRI scans, and 1228 CT scans from the TotalSegmentator dataset. A human-in-the-loop annotation workflow was employed, leveraging cross-modality transfer learning from an existing CT segmentation model to segment 40 anatomical structures. The annotation process began with a model based on transfer learning between CT and MR, which was iteratively refined based on manual corrections to predicted segmentations, until the entire training dataset was annotated.  The model’s performance was evaluated on MRI examinations obtained from the German National Cohort (NAKO) study (n=900) from the AMOS22 dataset (n=60) and from the TotalSegmentator-MRI test data (n=29). The Dice Similarity Coefficient (DSC) and Hausdorff Distance (HD) were used to assess segmentation quality, stratified by organ and scan type. The model and its weights will be open-sourced.
    
    \textbf{Results:} MRSegmentator demonstrated high accuracy for well-defined organs (lungs: DSC 0.96, heart: DSC 0.94) and organs with anatomic variability (liver: DSC 0.96, kidneys: DSC 0.95). Smaller structures showed lower accuracy (portal/splenic veins: DSC 0.64, adrenal glands: DSC 0.69). On external validation using NAKO data, mean DSC ranged from 0.85 ± 0.08 for T2-HASTE to 0.91 ± 0.05 for in-phase sequences. The model generalized well to CT, achieving mean DSC of 0.84 ± 0.11 on AMOS CT data.
    
    \textbf{Conclusion:} MRSegmentator accurately segments 40 anatomical structures in MRI across diverse datasets and imaging protocols, with additional generalizability to CT images. This open-source model will provide a valuable tool for automated multi-organ segmentation in medical imaging research. It can be downloaded from \url{https://github.com/hhaentze/MRSegmentator}.
}
\endabstract

\section{Introduction}\label{sec1}
Automated segmentation of anatomical structures in medical images enables precise organ volumetry, facilitates anatomical context for AI-based diagnosis, and supports quantitative imaging biomarker extraction \cite{gillies2016radiomics}. While recent advances in deep learning have led to robust CT segmentation models like TotalSegmentator, which segments 104 anatomical structures \cite{wasserthal2023totalsegmentator}, comparable tools for MRI segmentation remain limited, particularly for whole-body applications. 
Developing automated MRI segmentation poses distinct technical challenges compared to CT. MRI lacks standardized signal intensity values like Hounsfield units, exhibits variable tissue contrast across protocols, and often contains artifacts from motion or field inhomogeneities \cite{litjens2017survey}. The lower spatial resolution and often encountered anisotropic voxel sizes in MRI further complicate consistent segmentation. Despite these challenges, MRI segmentation tools are particularly valuable due to MRI's superior soft tissue contrast and lack of ionizing radiation, making it preferred for longitudinal studies and tissue characterization.

Current MRI segmentation solutions are predominantly organ-specific, focusing on individual structures like the kidneys, prostate, or spleen \cite{langner2019fully,kart2021deep,adams2022prostate158}. While these specialized models achieve high accuracy for their target organs, the need to run multiple models for different structures limits clinical workflow integration and adds computational overhead.  Recent work thus has proposed multi-organ approaches \cite{zhou2024mrannotator,zhuang2024mrisegmentator,d2024totalsegmentator}, but their generalizability across MRI sequences and external datasets remains unclear. The growing availability of large-scale MRI repositories, such as the UK Biobank imaging study, creates opportunities for developing more comprehensive segmentation tools that work across multiple protocols and anatomical regions.

To address this need, we have developed MRSegmentator, a nnU-Net based \cite{isensee2021nnu}, cross modality image segmentation model that segments 40 anatomical structures in both MRI and CT images. Our approach combines cross-modality learning from CT data \cite{ji2022amos} with a human-in-the-loop annotation workflow to accelerate image annotation and address MRI-specific challenges. We evaluate the model's performance across diverse MRI sequences using three external datasets and demonstrate its ability to handle anatomical variants. To facilitate research applications and further development, we have made MRSegmentator's code and trained weights publicly available at \url{https://github.com/hhaentze/MRSegmentator}.

\section{Materials \& Methods}\label{sec2}
This study was conducted in accordance with the Declaration of Helsinki and approved by the local ethics committee (EA4/062/20). Patient consent was waived due to the retrospective nature of the study.

\subsection{Datasets}
\begin{figure}[htbp]
    \centering
    \includegraphics[width=1\linewidth]{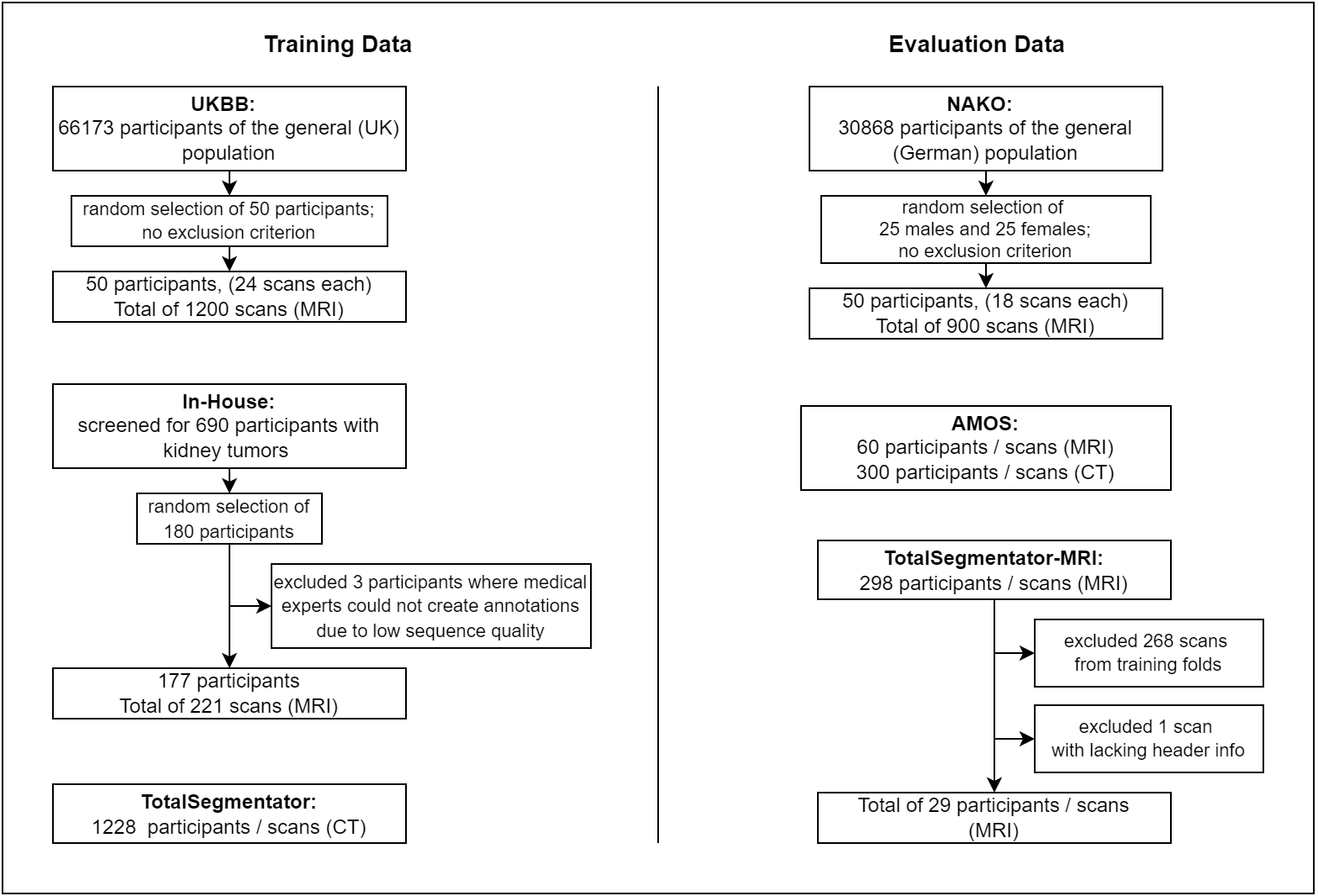}
    \caption{Study workflow with training and evaluation data. }
    \label{fig:data_workflow}
\end{figure}

Our study included six datasets of 3D MRI or CT images (Figure \ref{fig:data_workflow}). For training, we use data from the UK Biobank (UKBB), an in-house dataset, and the TotalSegmentator CT dataset. For testing, we use MRI scans from the National German Cohort (NAKO), the Multimodality Abdominal Multi-Organ Segmentation Challenge (AMOS22), and the TotalSegmentator MRI dataset. The datasets are described in more detail below.

\subsubsection{UK Biobank Dataset}
The UKBB is a large biomedical database containing genetic and health information from half a million UK participants \cite{littlejohns2020uk}. We accessed the MRI subset (datafield 20201-2, October 2023), which contains whole-body MRI scans from 69,571 participants. For our study, we selected 1,200 scans from 50 randomly selected participants. Each participant's MRI data is divided into six 3D sections from the shoulders to the knees, with in-phase (IN), out-of-phase (OPP), fat-only (F), and water-only (W) images obtained using the Dixon technique for each segment. 
This dataset is highly standardized, with consistent image size, voxel spacing, and subject positioning within each region. Each sequence consists of 44 to 72 axial slices spaced 3 to 4.5 mm along the z-axis. Access to the UKBB dataset for scientific research is available upon request at \url{https://www.ukbiobank.ac.uk/}.

\subsubsection{In-House Dataset}
We screened our in-house data for participants with kidney tumors or cysts. From the resulting 690 participants, we randomly selected 180. We also excluded scans from three participants due to poor image quality. Some participants had multiple sequence types, resulting in a final dataset of 221 axial abdominal MRI scans from 177 participants. Tumor size is less than 7 cm in 213 scans and up to 13 cm in the remaining eight scans. The dataset has an approximately equal distribution of T1, T2-weighted fat-saturated (T2fs), and post-contrast T1-weighted fat-saturated (T1fs) sequences. Images were acquired on different scanners with different signal intensity distributions and matrix sizes.
After acquisition, the sequences were exported and resized to a uniform voxel spacing of (1,1,1) mm, resulting in sequences consisting of 100 to 450 slices along the z-axis.

\subsubsection{TotalSegmentator Dataset}
To enable cross modality segmentation capabilities, we included the TotalSegmentator CT dataset, which consists of 1,228 CT examinations with a wide range of different pathologies, scanners and acquisition protocols \cite{wasserthal2023totalsegmentator}. From the 117 segmented structures, we selected a subset of 40 classes, consistent with the classes annotated for MR images (Figure \ref{fig:sample_seg}). Vertebrae, left lung lobe and right lung lobe are combined into common classes, as they are not reliable differentiable on out training MRI data, due to poor tissue contrast in the lungs and anisotropic voxel spacing. The dataset is available at \url{https://zenodo.org/records/10047292}. 

\subsubsection{NAKO Dataset}
The German National Cohort (NAKO) is a population-based prospective cohort study investigating the causes of the development of major chronic diseases in the German population \cite{de2014german}. The study includes whole-body MRI scans of 30,868 participants from the general population \cite{bamberg2015whole}. We obtained a subset of 900 MRI scans from 50 participants (25 women, 25 men). For each subject, the data included 1-weighted 3D VIBE two-point Dixon images (in phase (IN), opposed phase (OPP), water only (W), fat only(F)) and T2-weighted HASTE sequences (Table \ref{tab:data_composition}).
The Dixon sequences consist of four 3D section from the shoulders to the knees, and the T2 HASTE scans consist of two sections from the shoulder to the sacrum. We stitched the MRI stations for each participant and modality using a self-developed open-source utility (\url{https://github.com/ai-assisted-healthcare/AIAH_utility}) based on SimpleITK, resulting in 250 whole-body MRI sequences across contrast types. The stitched Dixon sequences have a matrix size of (320,260,316) with voxel spacing of (1.4,1.4,3.0) mm and the stitched T2-weighted HASTE sequences have a matrix size of (320,260,80) with voxel spacing of (1.4,1.4,6.0) mm.

\subsubsection{AMOS22 Dataset}
The Multimodality Abdominal Multi-Organ Segmentation Challenge (AMOS22) was held at the MICCAI conference in 2022 \cite{ji2022amos}. The accessible training and validation sections include 300 CT and 60 MRI images from multi-center, multi-vendor, multi-modality and multi-disease patients, each with voxel-level annotations of 15 abdominal organs. We excluded the classes 'prostate', as it is not part of our target classes, and 'bladder', which we believe to be incorrectly annotated in the AMOS scans. (Figure S1). The specific sequence and scanner types are not disclosed in the AMOS paper. The dataset is available here: \url{https://zenodo.org/records/7262581}.

\subsubsection{TotalSegmentator-MRI dataset}
TotalSegmentator-MRI \cite{d2024totalsegmentator} is a whole-body MRI segmentation model developed by D'Antonoli et al. and is the successor of TotalSegmentator for CT images. Given the similar names of the two models, the image modality will be specified when referring to both models. The openly available dataset of TotalSegmentator-MRI consists of 298 MRI scans, of which 30 images are marked as test data. We excluded one test scan (focused on the brain) due to incomplete header information. This dataset includes scans from different MRI scanners, covering multiple body regions, and including different sequence types in axial, sagittal, and coronal planes. It contains annotations for 59 anatomical structures; for our analysis, we selected the 40 structures that correspond to the segmentation targets of the MRSegmentator. The dataset is publicly available at \url{https://zenodo.org/doi/10.5281/zenodo.11367004}.

\subsection{Annotation Strategy}\label{sec_annostrat}
\begin{figure}
    \centering
    \includegraphics[width=1\linewidth]{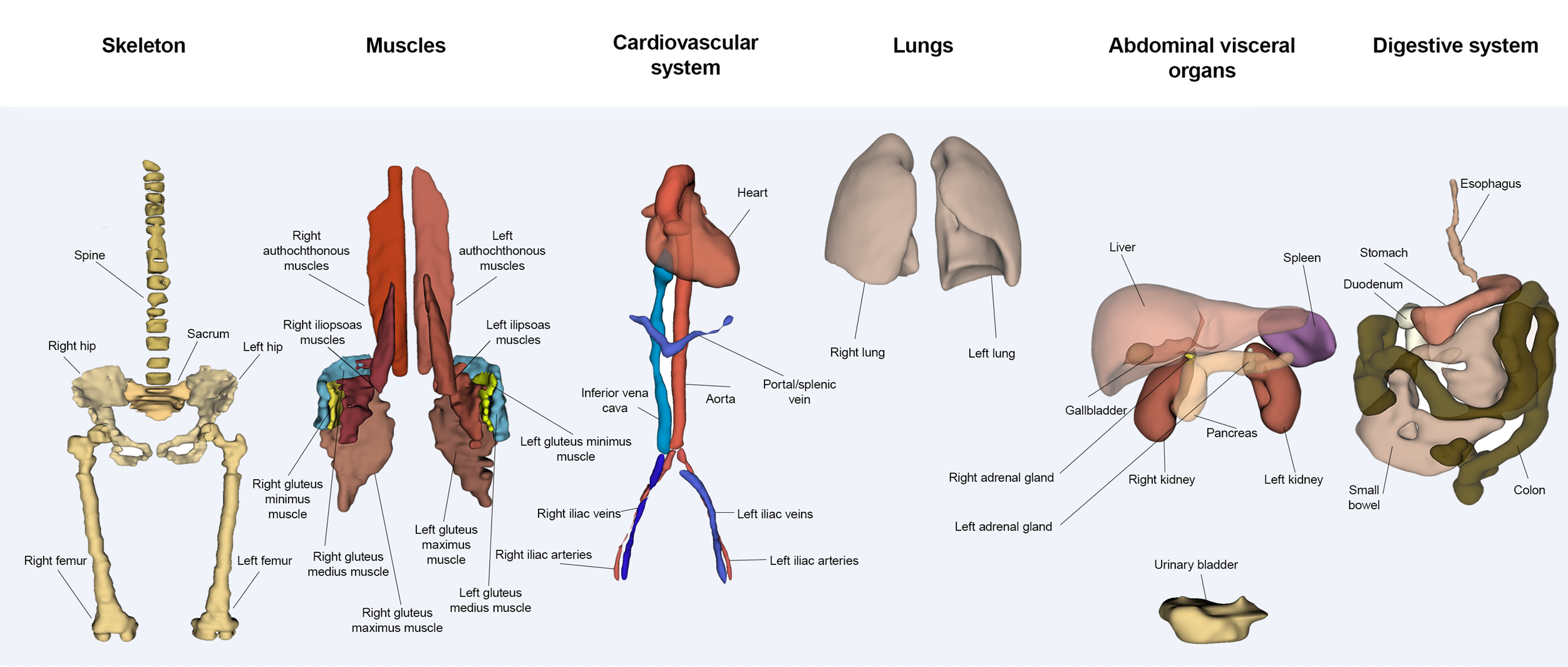}
    \caption{Sample segmentation output of MRSegmentator for all 40 classes. This includes spine, sacrum, hips, femurs, heart, aorta, inferior vena cava, portal/splenic vein, iliac arteries and veins, left and right lungs, liver, spleen, pancreas, gallbladder, stomach, duodenum, small bowel, colon, left and right kidneys, adrenal glands, urinary bladder, and muscles, specifically, gluteal muscles, autochthonous muscles, iliopsoas muscles. The model was trained on diverse datasets including UK Biobank, in-house clinical data, and CT scans, using a human-in-the-loop annotation approach. It demonstrates robust performance across various MRI sequences and can also segment CT images.}
    \label{fig:sample_seg}
\end{figure}

We developed a four-stage human-in-the-loop annotation workflow to create high-quality segmentations.
\begin{enumerate}
    \item Pre-segmentation: First, we generated initial segmentations by applying TotalSegmentator-CT to the MRI scans. To improve performance of the TotalSegmentator-CT model on MRI scans, we used preprocessing steps including intensity inversion and histogram equalization \cite{haentze2024improve}. For UK Biobank data, we segmented water-only sequences and propagated labels to the remaining Dixon sequences. While some structures like kidneys required only minor corrections, others such as muscles and bones needed complete re-annotation due to poor initial segmentation quality.
    \item Manual annotation: Three radiologists with one (LX)  and eight (LA, KB) years of experience in diagnostic radiology, refined and reviewed the pre-segmentations using MONAI Label \cite{diaz2024monai} and 3D Slicer \cite{fedorov20123d}). Overall, 40 different labels were created, which are detailed in Figure \ref{fig:sample_seg}. 
\item Model training: A nnU-Net \cite{isensee2021nnu} model was repeatedly trained, each time 50 new MRI sequences were annotated, enabling the generation of more refined labels, which were reviewed and refined again by the radiologists. This was repeated until the full training dataset of 1,200 UKBB scans and 221 in-house scans was annotated. Once the annotation process was complete, we trained the final nnU-Net with five-fold cross-validation on the fully annotated images, resulting in the final MRSegmentator model. This final training was performed using the 3d\_fullres\_no\_flipping configuration of nnU-Net V2 (\url{https://github.com/MIC-DKFZ/nnUNet}) with an increased batch size of eight. Other training parameters were kept at default values.  
    \item Test data annotation: After the MRSegmentator model was trained, the test data was manually annotated by the three radiologists.
\end{enumerate}

\begin{table}[h]
    \centering
    \caption{Data Composition}
    \label{tab:data_composition}
    \begin{threeparttable}
    \begin{tabular}{lccc}
     \multicolumn{4}{l}{\textbf{Training Data}} \\
      \hline
      \specialrule{0em}{0.3ex}{0.3ex} 
      \thead{} & \textbf{In-House} & \textbf{UKBB} & \textbf{TotalSegmentator} \\
      Nr. Participants & 171 (M:121, F:56)  & 50 (M:19, F: 31)  & 1228 (M: 716, F:510) \\
      \specialrule{0em}{0.3ex}{0.3ex} 
      Nr. Scans & 221(M:150, F:71)  & 1200 (M:456, F: 744) & 1228 (M: 716, F:510) \\
      \specialrule{0em}{0.3ex}{0.3ex} 
      Age [years] & 37 - 83 (median=62) & 40-69\textsuperscript{1} & 15 - 98 (median=65)\\
      \specialrule{0em}{0.3ex}{0.3ex} 
      Scanner Types & 1.5 and 3 Tesla MRI  & 1.5 Tesla MRI  & 20 different models\\
      \specialrule{0em}{0.3ex}{0.3ex}
      Sequences & \makecell{T1 (n=90) \\ T2fs (n=64) \\ T1fs  (n=67) } & \makecell{IN (n=300) \\ OPP (n=300) \\ W (n=300) \\ F (n=300)} & CT (n=1228)\\ 
      \\
      \multicolumn{4}{l}{\textbf{Test Data}} \\
      \hline
      \specialrule{0em}{0.3ex}{0.3ex} 
      \thead{} & \textbf{NAKO (MRI)} & \textbf{AMOS\textsuperscript{2}} & \textbf{TotalSegmentator MRI} \\
      Nr. Participants & 50 (M:25, F:25)  & \makecell{ MRI: 60 (M: 55, F: 45) \\ CT: 300 (M: 314, F: 186)} & 29 (M: 10, F:10, n.a.: 9) \\
      \specialrule{0em}{0.3ex}{0.3ex}
      Nr. Scans & 900 (M:450, F:450)  & \makecell{MRI: 60 (M: 55, F: 45)  \\ CT: 300 (M: 314, F: 186)}  & 29 (M: 10, F:10, n.a.: 9)\\
      \specialrule{0em}{0.3ex}{0.3ex}
      Age [years] & 26-69 (median=52.5) & \makecell{ MRI: 22-85 (median=50)  \\ CT: 14-94 (median=54)}  & 14-78 (median 60.5)\\
      \specialrule{0em}{0.3ex}{0.3ex}
      Scanner Types & 3 Tesla MR  & \makecell{MRI: 3 different models  \\ CT: 5 different models } & 13 different models\\
      \specialrule{0em}{0.3ex}{0.3ex}
      Sequences & \makecell{T1 GRE IN (n=200) \\ T1 GRE OPP (n=200) \\ T1 GRE W (n=200) \\ T1 GRE F (n=200) \\ T2 HASTE (n=100) } & \makecell{ MRI (n=60) \\ CT (n=300)} & MRI (n=29)\\
      \specialrule{0em}{0.1ex}{0.1ex}
      \hline
    \end{tabular}
    \end{threeparttable}
    \vspace{0.75em}
    
    \noindent
    \parbox{0.95\textwidth}{ 
        Table 1: Data composition of training and test data. (1) We did not have the social-demographics metadata of the UK Biobank data. Generally, the UK Biobank targets healthy individuals between the ages of 40 and 69. (2) Age and gender of the participants of the AMOS22 dataset are given for all 600 images, of which 360 are publicly available with manual annotations. Specific sequence and scanner types are not disclosed in the AMOS paper.
    }
\end{table}

\subsection{Statistical Analysis}
We assessed segmentation performance using three metrics: Dice Similarity Coefficient (DSC), 95th percentile Hausdorff Distance (HD), and a novel vessel consistency (VC) metric. DSC and HD were calculated per structure and sequence type, comparing model output against manual annotations. For thin, elongated structures like blood vessels, where DSC and HD can be misleading, we introduced VC as a complementary measure. VC is defined as the proportion of segmentations containing a single connected component for a given class, providing additional information about anatomical plausibility of vessel segmentations, putting emphasis on the continuity of the vessel segmentation.
To evaluate potential demographic effects on segmentation quality, we analyzed the relationship between DSC and participant characteristics in the NAKO dataset, which offered balanced demographics (25 men, 25 women). Sex-based differences in segmentation performance were assessed using independent two-sided t-tests, while age-related effects were evaluated using Pearson's correlation coefficient, both implemented in SciPy \cite{virtanen2020scipy}. With 50 participants (1:1 male-to-female ratio), our study achieved a power of 0.41 to detect medium-sized sex-based differences \cite{faul2007g}.
At the time of writing only one openly available whole body MRI segmentation algorithm exists (TotalSegmentator-MRI), therefore, for model comparison, we evaluated MRSegmentator against TotalSegmentator-MRI on matching anatomical structures across all test datasets. All metrics are reported as mean ± standard deviation unless otherwise specified. A p value < 0.05 was considered statistically significant.

\section{Results}

\begin{figure}[htbp]
    \centering
    \includegraphics[width=1\linewidth]{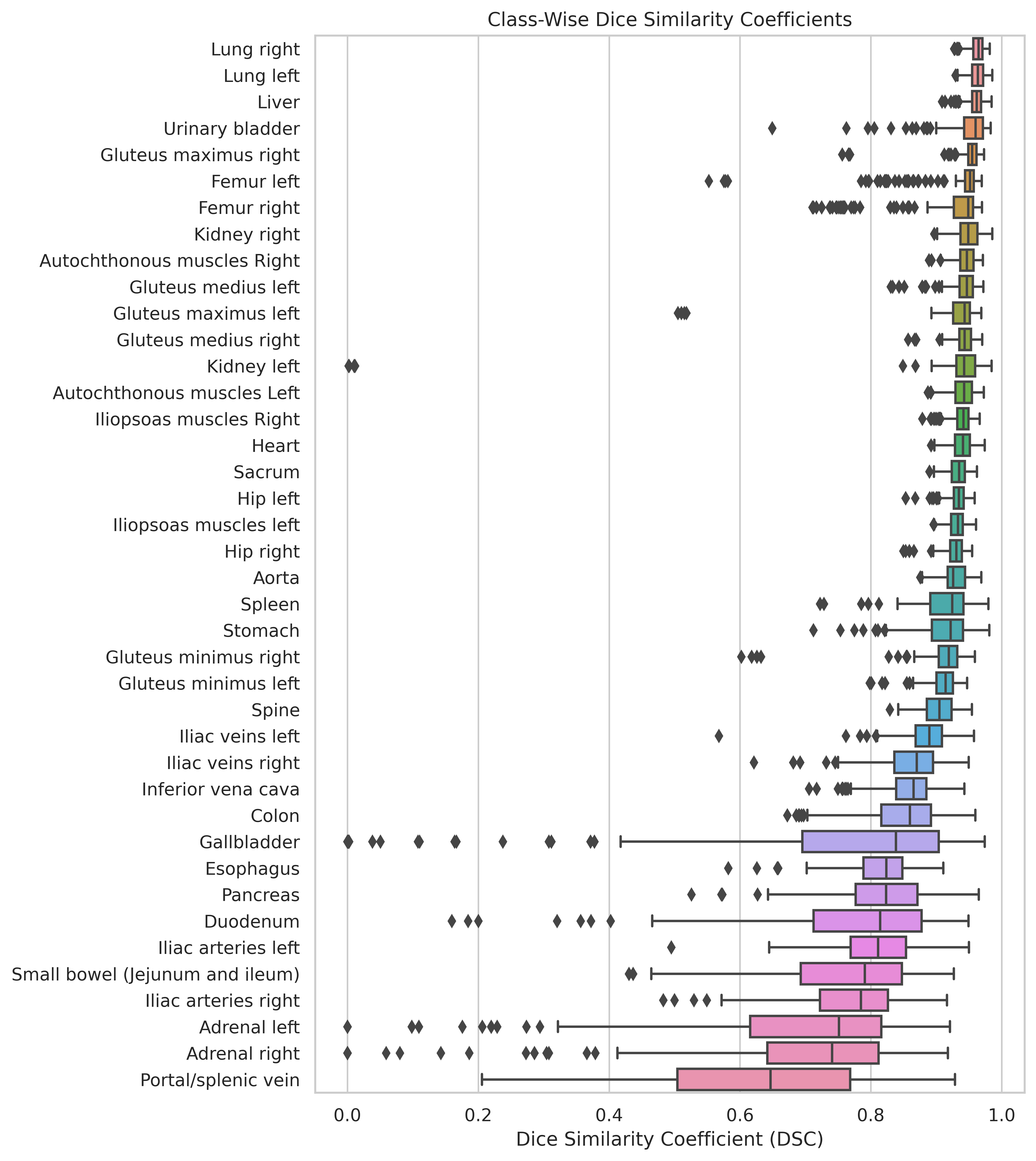}
    \caption{Class-wise Dice Similarity Coefficients (DSC) obtained by the MRSegmentator on the four GRE modalities of the NAKO dataset. The box plots show the distribution of DSC values for each of the 40 anatomical structures segmented by the model. The boxes represent the interquartile range (IQR), with the median DSC marked by the horizontal line within each box. The whiskers extend to the minimum and maximum values within 1.5 times the IQR, and any outliers beyond this range are represented by single points. The left kidney segmentations with a DSC of zero are false positives for a single patient who's left kidney was removed.}
    \label{fig:nako_dice}
\end{figure}

\subsection{Study Sample Characteristics}
We used three datasets for testing. The NAKO dataset consisted of 900 MRI scans from 50 patients (age range, 26-69 years; male-to-female ratio, 1:1). The AMOS22 dataset included 60 MRI scans and 300 CT scans from diverse patient populations (MRI age range, 22-85 years; CT age range, 14-95 years; male-to-female ratios, 1.2:1 and 1.7:1, respectively). TotalSegmentator-MRI test dataset contained 29 MRI scans (age range, 14-78 years; male-to-female ratio, 1:1, age and gender unknown for 9 scans).

\subsection{Segmentation Performance}
On NAKO data, MRSegmentator achieved averaged DSCs ranging from 0.85 ± 0.08 for T2-HASTE sequences to 0.91 ± 0.05 for T1-weighted Dixon in-phase sequences (Table \ref{tab:dice_comp}). Performance on the AMOS dataset yielded mean DSC scores of 0.79 ± 0.11 for MRI and 0.84 ± 0.11 for CT. The highest DSCs were observed for well-defined organs (lungs: 0.96, heart: 0.94, Figure \ref{fig:nako_dice}) and organs with anatomical variability (liver: 0.96, right kidney: 0.95, left kidney: 0.93). Small structures proved most challenging, particularly the portal/splenic vein (0.64) and adrenal glands (0.56 for AMOS).
Vessel Consistency analysis of NAKO Dixon examinations demonstrated high reliability for major vessels. The aorta and inferior vena cava were consistently segmented as single connected structures (VC: 100\% and 92\%, respectively; Table \ref{tab:vessel}). Iliac veins showed higher consistency (left: 0.94, right: 0.85) than arteries (left: 0.69, right: 0.59), while the portal/splenic vein typically appeared as multiple components (VC: 0.40).
Demographic analysis revealed superior segmentation quality in males (DSC = 0.89 ± 0.02) compared to females (DSC = 0.87 ± 0.02) in NAKO GRE sequences (p = 0.009), with the largest differences in adrenal glands ($\Delta$DSC = 0.13/0.10) and duodenum ($\Delta$DSC = 0.11). Participant age positively correlated with DSC (r = 0.37, p = 0.009; Figure S2).

\begin{table}
\centering
\caption{Vessel Consistency in External NAKO GRE Data}
\label{tab:vessel}
\begin{tabular}{lccc}
\hline
\specialrule{0em}{0.3ex}{0.3ex} 
\textbf{Vessel} & \textbf{VC} & \textbf{\makecell{Fraction of samples with\\more than one component}} & \textbf{\makecell{Average number of components\\if segmented incorrectly}} \\
\hline
\specialrule{0em}{0.3ex}{0.3ex} 
Aorta (n=200) & 1.00 & 0.00 & — \\
Inferior vena cava (n=200) & 0.92 & 0.08 & 2.06 ± 0.25 \\
Portal/splenic vein (n=200) & 0.40 & 0.60 & 2.60 ± 0.87 \\
Iliac arteries left (n=200) & 0.69 & 0.32 & 2.60 ± 0.85 \\
Iliac arteries right (n=200) & 0.59 & 0.41 & 2.54 ± 0.88 \\
Iliac veins left (n=200) & 0.94 & 0.06 & 2.18 ± 0.40 \\
Iliac veins right (n=200) & 0.85 & 0.15 & 2.13 ± 0.43 \\
\hline
\end{tabular}

\vspace{0.75em}
\parbox{0.95\textwidth}{ 
    Table 2: Vessel consistency (VC) refers to the proportion of segmentations where a given vessel class is represented by a single connected component. A higher VC indicates that the model effectively treats the vessel as a unified entity. MRSegmentator did not fail to segment any structure; therefore, the fraction of samples with multiple components is directly related to the inverse of the VC. The final column presents the average number of components detected when multiple components were identified. For example, segmentations of the portal/splenic vein exhibit a VC of 40\%, indicating that, in 60\% of cases, multiple components are detected. The average number of components observed in these cases is 2.6, suggesting a high degree of fragmentation in the segmentations for this vessel class.
}
\end{table}

\subsection{Comparison with TotalSegmentator-MRI}
MRSegmentator demonstrated superior performance compared to TotalSegmentator-MRI across NAKO and AMOS datasets, achieving higher mean DSC (NAKO: 0.91 ± 0.05 vs 0.83 ± 0.07; AMOS: 0.79 ± 0.11 vs 0.75 ± 0.12) and lower HD values (NAKO: 7.5 ± 14.8 mm vs 15.1 ± 20.3 mm; AMOS: 8.4 ± 7.1 mm vs 10.0 ± 8.7 mm) for all classes. The performance difference was most pronounced in abdominal organs (liver: 0.96 vs 0.89, spleen: 0.91 vs 0.84, pancreas: 0.82 vs 0.73) and blood vessels (aorta: 0.93 vs 0.85, inferior vena cava: 0.86 vs 0.78).
On the TotalSegmentator-MRI test set, our model achieved comparable overall performance (DSC: 0.74 ± 0.21 vs 0.76 ± 0.21) but showed distinct strengths in different anatomical regions. MRSegmentator excelled in abdominal organ and vessel segmentation, while TotalSegmentator-MRI demonstrated better performance in musculoskeletal structures.

\begin{table}[htbp]
\centering
\caption{Comparison of MRSegmentator and TotalSegmentator-MRI across all datasets}
\label{tab:dice_comp}
\begin{tabular}{lcccc}
\specialrule{0em}{0.3ex}{0.3ex} 
\hline
\textbf{Dataset} & \multicolumn{2}{c}{\textbf{DSC}} & \multicolumn{2}{c}{\textbf{HD [mm]}} \\
                 & MRSeg\textsuperscript{1} & TotalSegMRI\textsuperscript{2} & MRSeg\textsuperscript{1} & TotalSegMRI\textsuperscript{2} \\
\hline
\specialrule{0em}{0.3ex}{0.3ex} 
NAKO in-phase (40 labels) & \textbf{0.91 $\pm$ 0.05} & 0.83 $\pm$ 0.07 & \textbf{7.5 $\pm$ 14.8} & 15.1 $\pm$ 20.3 \\
NAKO opposed-phase (40 labels) & \textbf{0.88 $\pm$ 0.05} & 0.82 $\pm$ 0.06 & \textbf{8.4 $\pm$ 15.0} & 12.7 $\pm$ 17.4 \\
NAKO water only (40 labels) & \textbf{0.87 $\pm$ 0.05} & 0.81 $\pm$ 0.06 & \textbf{8.5 $\pm$ 15.0} & 13.1 $\pm$ 16.3 \\
NAKO fat only (40 labels) & \textbf{0.88 $\pm$ 0.06} & 0.80 $\pm$ 0.07 & \textbf{8.3 $\pm$ 14.8} & 15.6 $\pm$ 20.6 \\
TotalSegmentator-MRI data (40 labels) & 0.74 $\pm$ 0.21 & \textbf{0.76 $\pm$ 0.21} & 16.1 $\pm$ 21.7 & \textbf{16.7 $\pm$ 19.7} \\
\specialrule{0em}{0.1ex}{0.1ex}
\hline
\specialrule{0em}{0.3ex}{0.3ex}
NAKO T2 HASTE (24 labels) & \textbf{0.85 $\pm$ 0.08} & 0.75 $\pm$ 0.10 & \textbf{8.5 $\pm$ 7.0} & 14.1 $\pm$ 9.5 \\
NAKO in-phase (24 labels)\textsuperscript{3} & \textbf{0.90 $\pm$ 0.06} & 0.82 $\pm$ 0.08 & \textbf{6.2 $\pm$ 7.2} & 11.3 $\pm$ 8.1 \\
\specialrule{0em}{0.1ex}{0.1ex}
\hline
\specialrule{0em}{0.3ex}{0.3ex} 
Amos MRI (13 labels) & \textbf{0.79 $\pm$ 0.11} & 0.75 $\pm$ 0.12 & \textbf{8.4 $\pm$ 7.1} & 10.0 $\pm$ 8.7 \\
Amos CT (13 labels)\textsuperscript{4} & 0.84 $\pm$ 0.11 & 0.84 $\pm$ 0.11 & \textbf{7.9 $\pm$ 11.3} & 8.1 $\pm$ 10.5 \\
NAKO in-phase (13 labels)\textsuperscript{3} & \textbf{0.88 $\pm$ 0.09} & 0.78 $\pm$ 0.12 & \textbf{4.3 $\pm$ 3.6} & 10.1 $\pm$ 6.0 \\
\hline
\end{tabular}
\vspace{0.75em}

\noindent
\parbox{0.95\textwidth}{
    Table 3: We compared (1) MRSegmentator and (2) TotalSegmentator-MRI on different datasets and reported the Dice Similiarity Coefficient (DSC),  where larger values
    indicate better performance, and Hausdorff-distance (HD),  where smaller values indicate better performance. Both metrics are accompanied by their standard deviation. (3) The NAKO T2 and the AMOS data have fewer annotated classes. To allow a fair comparison to the NAKO GRE sequences we additionally report the results for the classes-subset on the in-phase sequences, as a representative for the GRE scans. (4) For the AMOS CT dataset we use TotalSegmentator-CT instead of TotalSegmentator-MRI as a baseline model.
}
\end{table}

\subsection{Failure cases and performance on pathological images}
\begin{figure}
    \centering
    \includegraphics[width=1\linewidth]{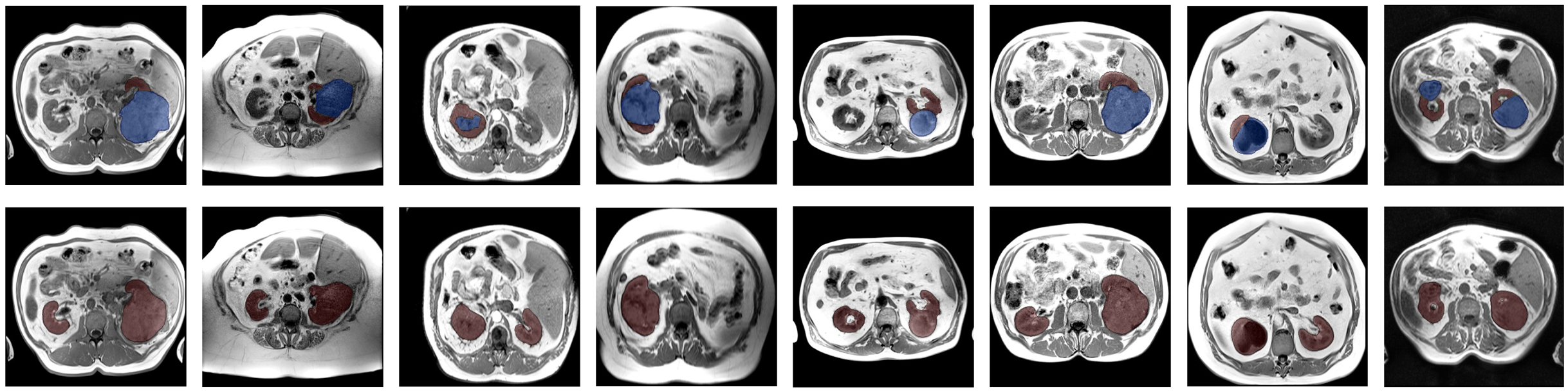}
    \caption{The first row depicts manually annotated kidneys (red) and tumors (blue) in eight different MRI scans. The second row shows the corresponding kidney segmentations generated by MRSegmentator. The model accurately localizes and segments the kidneys even in the presence of large tumors, demonstrating its robustness in handling pathological cases. In the fourth sample, the missing right kidney due to a previous nephrectomy is correctly not segmented by the model. (All images were taken from validation folds and not used for training in this analysis specifically.)}
    \label{fig:kidney_scans}
\end{figure}

Despite robust overall performance, we could identify specific failure patterns. Left-right confusion occurred occasionally in the pelvic region, evidenced by large HDs for femurs (left: 37.8 mm, right: 18.9 mm) and left iliopsoas muscle (left: 19.6 mm, right 6.0 mm). In the kidney tumor subset, MRSegmentator accurately segmented kidneys with tumors larger than 7 cm (n=8, Figure \ref{fig:kidney_scans}), though some oversegmentation occurred with irregular tumor borders. The model showed inconsistent performance in post-nephrectomy cases, correctly identifying single kidneys in validation data but occasionally misclassifying colon as kidney in test data.

\section{Discussion}
In this study, we developed MRSegmentator, a deep learning model that segments 40 anatomical structures across MRI sequences and CT images using a unified architecture. By combining cross-modality learning with human-in-the-loop annotation, we achieved robust performance across diverse imaging protocols and patient populations. The model demonstrated particular strength in segmenting parenchymal organs while maintaining acceptable accuracy for more challenging structures such as small vessels and glands.

Comparison with existing methods shows that MRSegmentator achieves competitive performance against specialized single-organ models. Our spleen and liver segmentation accuracy (DSC: 0.95, 0.96) matches dedicated models (0.96, 0.95) \cite{sharbatdaran2022deep,hossain2023deep},  while multi-organ performance for abdominal structures performance (DSC for liver, spleen, kidneys, pancreas: 0.96, 0.91, 0.94, 0.82 for NAKO MRI; 0.96, 0.95, 0.95, 0.81 for AMOS MRI) approaches that of recent organ-specific approaches, such as reported by Kart et al. (0.98, 0.96, 0.98, 0.89) \cite{kart2021deep}. Notably, MRSegmentator's ability to process both MRI and CT images with a single model (CT DSC: 0.84), achieving segmentation quality on par with TotalSegmentator-CT, the current state of the art in multi-organ segmentation in CT, offers practical advantages over modality-specific solutions, potentially streamlining clinical and research workflows. This versatility may outweigh slightly inferior performance compared to specialized models.

To address the limitations of the voxel-based metrics DSC and HD, we introduced the vessel consistency metric, which revealed that large vessels such as the aorta are consistently segmented as single structures (VC: 1.0), whereas smaller vessels such as the portal/splenic vein often appear fragmented (VC: 0.40). This metric provides insights not captured by traditional DSC measurements, particularly relevant for elongated structures. We observed gender differences in performance, with higher accuracy in male subjects ($\Delta$DSC = 0.02, p = 0.009), likely reflecting anatomical differences in muscle mass and organ positioning rather than technical limitations. The model showed robust performance on pathological cases, successfully segmenting kidneys with large tumors, while occasionally struggling with postoperative anatomy.

MRSegmentator will allow researchers to obtain biomarkers relevant for various research questions and clinical tasks. For instance, total kidney volume has been shown to correlate with glomerular filtration rate, a key indicator in polycystic kidney disease \cite{jo2017correlations}. Additionally, iliac artery tortuosity may serve as a predictor of biological age \cite{mach2021iliofemoral}, while the fat fraction within the autochthonous muscles can assist in stratifying the risk of incidental, non-traumatic vertebral fractures in the lower thoracic spine among elderly patients \cite{backhauss2023fatty}.

Our study has limitations. The human in the loop may have introduced annotation bias, however strong performance on fully independent external datasets suggests minimal impact. The UK Biobank training data, while numerous in scans (1,200), represents only 50 unique participants, potentially limiting anatomical variety. This limitation is partially mitigated by including diverse in-house and TotalSegmentator-CT data. Additionally, the observed gender-based performance differences highlight the need for more balanced training datasets.

\section{Conclusion}
MRSegmentator represents an advancement in multi-modal, medical image segmentation, demonstrating strong performance and generalizability across diverse datasets. The capability of segmenting CT and MR images, makes it a valuable tool for researchers and clinicians. We will continue to work on MRSegmentator by focusing on expanding the range of supported anatomical structures and pathological conditions while maintaining the model's cross-modality capabilities.

\section*{Acknowledgements}
This research has been conducted using the UK Biobank Resource under Application Number 105529. This project was conducted with data from the German National Cohort (NAKO) (www.nako.de ). The NAKO is funded by the Federal Ministry of Education and Research (BMBF) [project funding reference numbers: 01ER1301A/B/C, 01ER1511D and 01ER1801A/B/C/D], federal states of Germany and the Helmholtz Association, the participating universities and the institutes of the Leibniz Association. We thank all participants who took part in the NAKO study and the staff of this research initiative. Much of the computation resources required for this research was performed on computational hardware generously provided by the Charité HPC cluster (\url{https://www.charite.de/en/research/research_support_services/research_infrastructure/science_it/#c30646061}). This work was in large parts funded by the Wilhelm Sander Foundation. Funded by the European Union. Views and opinions expressed are however those of the author(s) only and do not necessarily reflect those of the European Union or European Health and Digital Executive Agency (HADEA). Neither the European Union nor the granting authority can be held responsible for them.

\begin{figure}[H]
\includegraphics[width=0.3\linewidth, right]{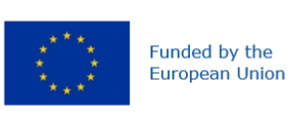}
\label{fig:eu_funding}
\end{figure}

\newpage

\bibliographystyle{unsrtnat}
\bibliography{references.bib}

\newpage
\section{Appendix}\label{sec6}

\renewcommand{\thefigure}{A\arabic{figure}} 
\renewcommand{\thetable}{A\arabic{table}} 
\setcounter{figure}{0} 
\setcounter{table}{0} 

\begin{table}[htbp]
\centering
\caption{Overview of different body regions and organs, supported by MRSegmentator}
\label{tab:classes}
\begin{tabular}{|l|l|}
\hline
\textbf{Body Region} & \textbf{Organs} \\ \hline
\multirow{3}{*}{Chest} & Heart \\
 & Left and right lungs \\
 & Esophagus \\ \hline
\multirow{8}{*}{Gastrointestinal tract} & Liver \\
 & Spleen \\
 & Pancreas \\
 & Gallbladder \\
 & Stomach \\
 & Duodenum \\
 & Small bowel (Jejunum and ileum) \\
 & Colon \\ \hline
\multirow{3}{*}{Retroperitoneum} & Left and right kidneys \\
 & Left and right adrenal glands \\
 & Urinary bladder \\ \hline
\multirow{9}{*}{Musculoskeletal} & Spine \\
 & Sacrum \\
 & Left and right hip \\
 & Left and right femurs \\
 & Left and right gluteus maximus muscles \\
 & Left and right gluteus medius muscles \\
 & Left and right gluteus minimus muscles \\
 & Left and right autochthonous muscles \\
 & Left and right iliopsoas muscles \\ \hline
\multirow{5}{*}{Vessels} & Aorta \\
 & Inferior vena cava \\
 & Portal/splenic vein \\
 & Left and right iliac arteries \\
 & Left and right iliac veins \\ \hline
\end{tabular}
\end{table}

\begin{table}[htbp]
\centering
\caption{Comparison of Dice similarity coefficient for different test datasets}
\label{tab:classwise_results}
\begin{tabular}{|c|c|c|c|c|c|}
\hline
\textbf{Structure} & \textbf{NAKO GRE} & \textbf{NAKO T2} & \textbf{AMOS22 (MR)} & \textbf{AMOS22 (CT)} & \textbf{TotSeg-MRI\textsuperscript{*}} \\ \hline
Lung left & 0.96 & 0.96 & ... & ... & 0.92 \\ \hline
Lung right & 0.96 & 0.96 & ... & ... & 0.81 \\ \hline
Heart & 0.94 & 0.92 & ... & ... & 0.81 \\ \hline
Esophagus & 0.81 & 0.79 & 0.66 & 0.80 & 0.61 \\ \hline
Liver & 0.96 & 0.95 & 0.96 & 0.96 & 0.82 \\ \hline
Spleen & 0.91 & 0.91 & 0.95 & 0.95 & 0.88 \\ \hline
Pancreas & 0.82 & 0.80 & 0.81 & 0.79 & 0.45 \\ \hline
Gallbladder & 0.74 & 0.75 & 0.74 & 0.80 & 0.83 \\ \hline
Stomach & 0.91 & 0.84 & 0.87 & 0.89 & 0.86 \\ \hline
Duodenum & 0.77 & 0.43 & 0.58 & 0.67 & 0.70 \\ \hline
Small bowel (Jejunum \& Ileum) & 0.76 & 0.57 & ... & ... & 0.78 \\ \hline
Colon & 0.85 & 0.82 & ... & ... & 0.82 \\ \hline
Kidney left & 0.93 & 0.95 & 0.95 & 0.94 & 0.77 \\ \hline
Kidney right & 0.95 & 0.95 & 0.95 & 0.94 & 0.88 \\ \hline
Adrenal left & 0.69 & 0.69 & 0.56 & 0.69 & 0.66 \\ \hline
Adrenal right & 0.69 & 0.72 & 0.56 & 0.70 & 0.56 \\ \hline
Urinary bladder & 0.95 & ... & ... & ... & 0.89 \\ \hline
Spine & 0.90 & 0.93 & ... & ... & 0.69 \\ \hline
Sacrum & 0.93 & ... & ... & ... & 0.78 \\ \hline
Hip left & 0.93 & ... & ... & ... & 0.65 \\ \hline
Hip right & 0.93 & ... & ... & ... & 0.58 \\ \hline
Femur left & 0.93 & ... & ... & ... & 0.80 \\ \hline
Femur right & 0.92 & ... & ... & ... & 0.85 \\ \hline
Gluteus maximus left & 0.93 & ... & ... & ... & 0.83 \\ \hline
Gluteus maximus right & 0.95 & ... & ... & ... & 0.73 \\ \hline
Gluteus medius left & 0.94 & ... & ... & ... & 0.85 \\ \hline
Gluteus medius right & 0.94 & ... & ... & ... & 0.75 \\ \hline
Gluteus minimus left & 0.91 & ... & ... & ... & 0.85 \\ \hline
Gluteus minimus right & 0.91 & ... & ... & ... & 0.83 \\ \hline
Autochthonous muscles feft & 0.94 & 0.95 & ... & ... & 0.70 \\ \hline
Autochthonous muscles right & 0.95 & 0.95 & ... & ... & 0.67 \\ \hline
Iliopsoas muscles left & 0.93 & 0.95 & ... & ... & 0.53 \\ \hline
Iliopsoas muscles right & 0.94 & 0.96 & ... & ... & 0.54 \\ \hline
Aorta & 0.93 & 0.94 & 0.90 & 0.93 & 0.71 \\ \hline
Inferior vena cava & 0.86 & 0.89 & 0.83 & 0.84 & 0.66 \\ \hline
Portal/splenic vein & 0.64 & 0.78 & ... & ... & 0.65 \\ \hline
Iliac arteries left & 0.81 & ... & ... & ... & 0.71 \\ \hline
Iliac arteries right & 0.77 & ... & ... & ... & 0.66 \\ \hline
Iliac veins left & 0.88 & ... & ... & ... & 0.74 \\ \hline
Iliac veins right & 0.86 & ... & ... & ... & 0.74 \\ \hline
\textbf{Mean} & \textbf{0.88} & \textbf{0.85} & \textbf{0.79} & \textbf{0.84} & \textbf{0.74} \\ \hline
\end{tabular}
\vspace{0.75em}
    
    \noindent
    \parbox{0.95\textwidth}{ 
        Table A2: Class-wise Dice similarity coefficients of MRSegmentator for NAKO, AMOS and \textsuperscript{*}TotalSegmentator-MRI test datasets.. Results for GRE scans are averaged across in-phase, opposed-phase, water-only, and fat-only sequences. For T2-HASTE scans, which cover from the lungs to the sacrum, results are reported only for anatomical structures fully captured within this field of view.
    }
\end{table}

\begin{table}[htbp]
\centering
\caption{Comparison of Hausdorff distance for different test datasets}
\label{tab:classwise_results}
\begin{tabular}{|c|c|c|c|c|c|}
\hline
\textbf{Structure} & \textbf{NAKO GRE} & \textbf{NAKO T2} & \textbf{AMOS22 (MR)} & \textbf{AMOS22 (CT)} & \textbf{TotSeg-MRI\textsuperscript{*}} \\ \hline
Lung left & 6.1 & 6.2 & ... & ... & 7.2 \\ \hline
Lung right & 6.0 & 7.3 & ... & ... & 8.7 \\ \hline
Heart & 4.7 & 6.3 & ... & ... & 12.1 \\ \hline
Esophagus & 4.2 & 7.3 & 12.4 & 9.0 & 17.2 \\ \hline
Liver & 4.9 & 7.1 & 4.7 & 6.1 & 25.8 \\ \hline
Spleen & 5.0 & 6.3 & 3.4 & 4.6 & 8.7 \\ \hline
Pancreas & 6.9 & 14.4 & 7.9 & 10.2 & 38.7 \\ \hline
Gallbladder & 5.3 & 7.0 & 7.8 & 9.1 & 13.5 \\ \hline
Stomach & 6.2 & 13.3 & 9.3 & 13.3 & 15.4 \\ \hline
Duodenum & 9.4 & 29.7 & 18.2 & 15.6 & 12.9 \\ \hline
Small bowel (Jejunum \& Ileum) & 15.5 & 31.3 & ... & ... & 29.3 \\ \hline
Colon & 12.9 & 16.8 & ... & ... & 11.2 \\ \hline
Kidney left & 4.1 & 3.9 & 3.5 & 3.5 & 16.4 \\ \hline
Kidney right & 3.8 & 3.2 & 3.9 & 4.3 & 6.8 \\ \hline
Adrenal left & 4.7 & 5.1 & 8.8 & 6.6 & 4.6 \\ \hline
Adrenal right & 3.8 & 4.1 & 7.8 & 6.6 & 5.7 \\ \hline
Urinary bladder & 3.4 & ... & ... & ... & 3.9 \\ \hline
Spine & 3.1 & 2.7 & ... & ... & 13.9 \\ \hline
Sacrum & 2.8 & ... & ... & ... & 6.8 \\ \hline
Hip left & 3.9 & ... & ... & ... & 17.6 \\ \hline
Hip right & 4.6 & ... & ... & ... & 17.4 \\ \hline
Femur left & 37.8 & ... & ... & ... & 16.0 \\ \hline
Femur right & 18.9 & ... & ... & ... & 5.7 \\ \hline
Gluteus maximus left & 5.2 & ... & ... & ... & 34.8 \\ \hline
Gluteus maximus right & 12.9 & ... & ... & ... & 36.2 \\ \hline
Gluteus medius left & 7.5 & ... & ... & ... & 3.6 \\ \hline
Gluteus medius right & 7.6 & ... & ... & ... & 4.2 \\ \hline
Gluteus minimus left & 7.8 & ... & ... & ... & 4.6 \\ \hline
Gluteus minimus right & 10.8 & ... & ... & ... & 5.5 \\ \hline
Autochthonous muscles feft & 8.8 & 7.7 & ... & ... & 19.6 \\ \hline
Autochthonous muscles right & 6.9 & 6.7 & ... & ... & 21.9 \\ \hline
Iliopsoas muscles left & 19.6 & 1.8 & ... & ... & 24.9 \\ \hline
Iliopsoas muscles right & 6.0 & 1.7 & ... & ... & 28.3 \\ \hline
Aorta & 2.7 & 2.5 & 15.6 & 13.8 & 21.4 \\ \hline
Inferior vena cava & 4.4 & 4.4 & 6.3 & 8.6 & 16.5 \\ \hline
Portal/splenic vein & 13.4 & 7.7 & ... & ... & 17.4 \\ \hline
Iliac arteries left & 15.1 & ... & ... & ... & 27.7 \\ \hline
Iliac arteries right & 6.5 & ... & ... & ... & 22.3 \\ \hline
Iliac veins left & 9.8 & ... & ... & ... & 19.4 \\ \hline
Iliac veins right & 4.8 & ... & ... & ... & 19.5 \\ \hline
\textbf{Mean} & \textbf{8.2} & \textbf{8.5} & \textbf{8.4} & \textbf{8.6} & \textbf{16.1} \\ \hline
\end{tabular}
\vspace{0.75em}
    
    \noindent
    \parbox{0.95\textwidth}{ 
        Table A3: Class-wise Hausdorff distances of MRSegmentator for NAKO, AMOS and \textsuperscript{*}TotalSegmentator-MRI test datasets. Results for GRE scans are averaged across in-phase, opposed-phase, water-only, and fat-only sequences. For T2-HASTE scans, which cover from the lungs to the sacrum, results are reported only for anatomical structures fully captured within this field of view.
    }
\end{table}

\begin{figure}[htbp]
    \centering
    \includegraphics[width=0.6\linewidth]{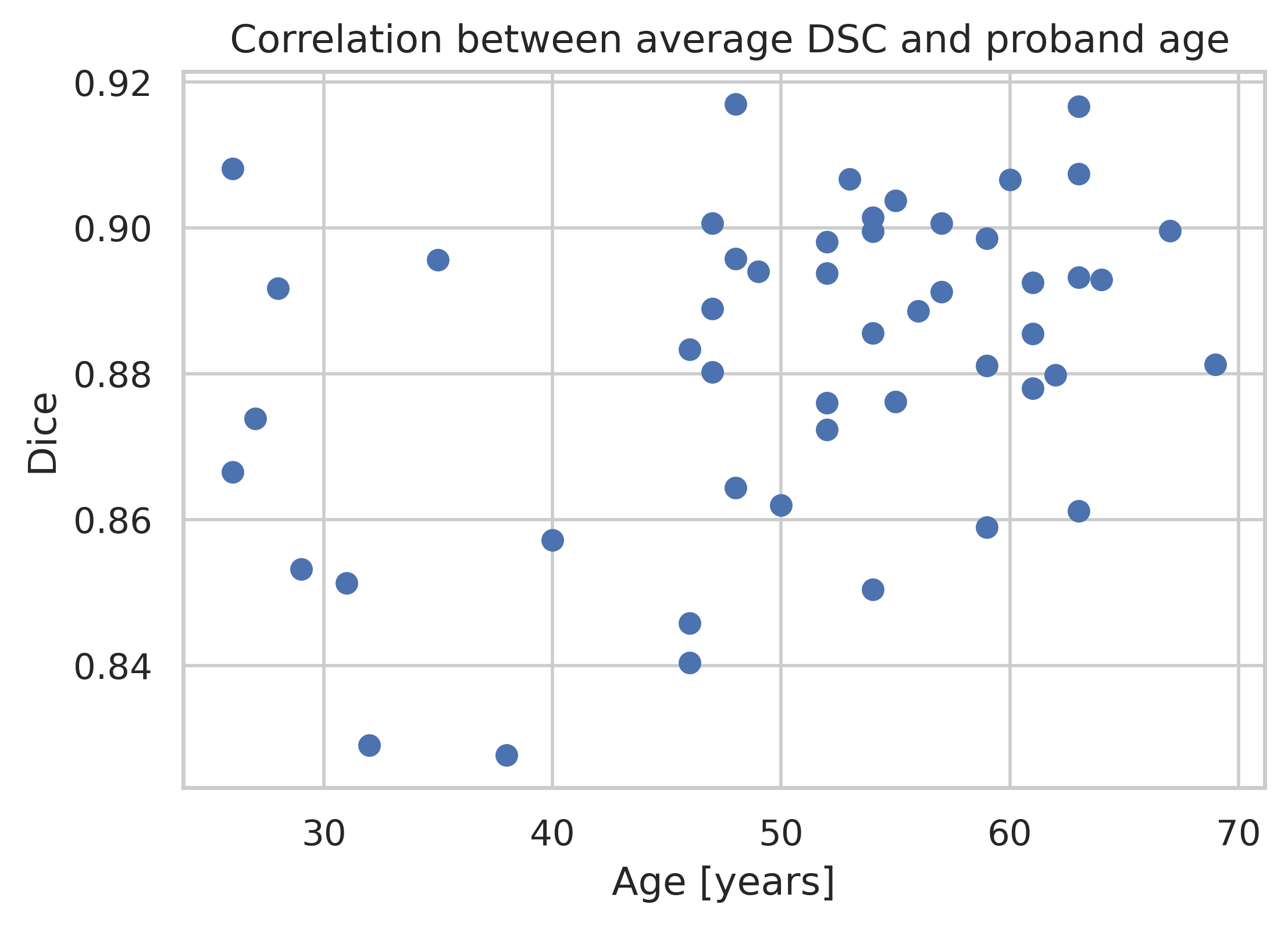}
    \caption{The Dice similarity coefficient correlates positively with participant age (r=0.37, p=0.009).}
    \label{fig:age_correlation}
\end{figure}

\begin{figure}[htbp]
    \centering
    \includegraphics[width=0.9\linewidth]{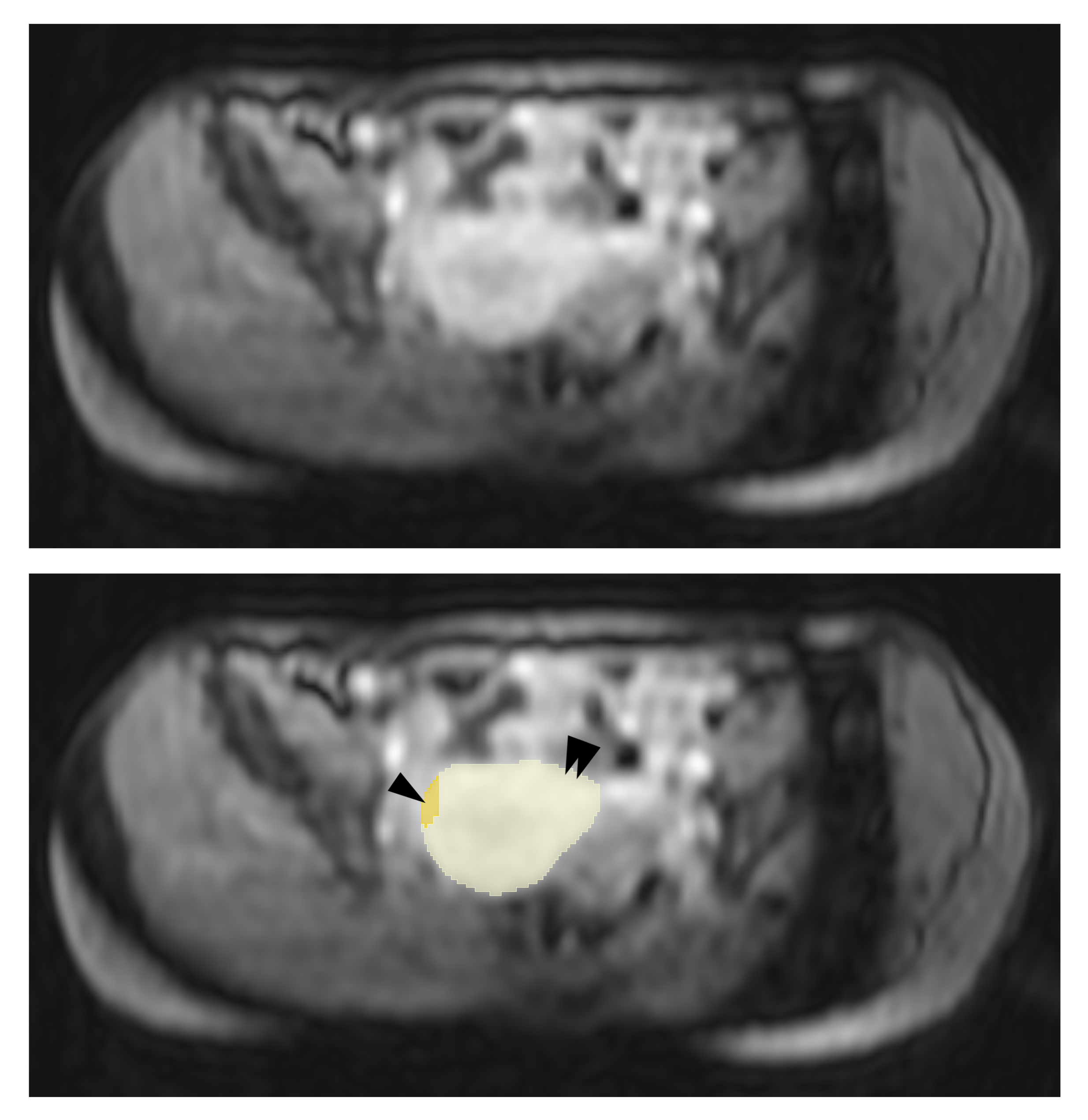}
    \caption{The AMOS22 dataset contains only two MRI examinations with bladder annotations. Of these two annotations, one incorrectly combines the bladder (yellow) with the uterus (beige). Consequently, we exclude the label "bladder" from our evaluation of the AMOS22 dataset.}
    \label{fig:wrong_annotation}
\end{figure}

\end{document}